\documentstyle[aps,epsf]{revtex}  

\begin{document}        

\baselineskip 14pt
\title{Lattice calculation of matrix elements relevant for $\Delta I=1/2$ 
rule and $\varepsilon '/\varepsilon$.}
\author{D. Pekurovsky, G. Kilcup}
\address{Department of Physics, The Ohio State University, Columbus, 
Ohio 43210, USA}
%
\maketitle              

\begin{abstract}        
We have gained enough statistical precision to distinguish signal from 
noise in matrix elements
of all operators relevant for the $\Delta I=1/2$ rule in kaon decays
and for the direct CP violation parameter $\varepsilon '$.
We confirm significant enhancement of $\Delta I=1/2$ transitions
observed in experiments, although a few large systematic 
uncertainties remain in our predictions: higher-order chiral
corrections and lattice spacing dependence.
The estimate of $\varepsilon '/\varepsilon$ is further complicated
by the problem of matching lattice and continuum operators.
\end{abstract}   	

\section{Introduction}               

One of the poorly understood effects in low-energy phenomenology is
the so-called $\Delta I=1/2$ rule. Namely, kaon decays proceed
at much higher rate through the $\Delta I=1/2$ channel than through
the $\Delta I=3/2$ one. In particular, for decays to two pions
the following relationship between the amplitudes is experimentally 
observed:
\begin{equation}
\label{22}
\frac{A(K\to (\pi\pi )_{I=0})}{A(K\to (\pi\pi )_{I=2})} \equiv 
\frac{A_0}{A_2}= 22\,.
\end{equation}

In the Standard Model, the electro-weak interactions and the 
short-distance part of strong interactions are not enough to explain 
this predominance of $\Delta I=1/2$ transitions. The bulk of the
ratio in Eq.~\ref{22} is attributed to the long-distance part
of strong interactions. The long-distance effects 
are contained in the matrix elements (MEs) of the basis operators
of the weak effective theory
between hadron states. These MEs have remained mostly unknown due to 
their non-perturbative character. We compute them on the 
lattice with statistics more than sufficient to distinguish signal 
from noise for the $\Delta I=1/2$ amplitude, thus allowing to 
check the prediction of the Standard Model and QCD against the 
experiment. 

In a related effort, we address 
the direct CP-violating parameter $\varepsilon '$, defined in
\begin{equation}
\varepsilon ' = 
\frac{\langle\pi^+\pi^-|K_2\rangle}{\langle\pi^+\pi^-|K_1\rangle} \, ,
\end{equation}
where $K_1$ ($K_2$) are CP-even (odd) strong interaction eigenstates.
Our goal is to predict the value of $\varepsilon '$ in the Standard
Model in order to see if it is different from the Superweak theory and 
to compare it with experiment. As for the experimental situation, 
Fermilab's KTEV group has recently announced their latest result 
of \mbox{Re$(\varepsilon '/\varepsilon) = (28 \pm 4.1) \cdot 10^{-4}$,}
which is consistent with the old result from CERN's NA48 experiment
$((23 \pm 7) \cdot 10^{-4}$),
although hardly consistent with the Fermilab's previous result of 
$(7.4 \pm 5.9) \cdot 10^{-4}$. The $\varepsilon '/\varepsilon$ value will 
be even more refined soon by including more data from Fermilab 
and by complementing it with results from two other high-precision 
experiments at CERN and $\phi$ factory at Frascati. The theory is lagging
behind these experiments, mostly due to the lack of knowledge of MEs 
containing long-distance dynamics.

\section{Simulation details}

We work within the effective theory obtained from the Standard
Model by integrating out the $W$ boson and $t$, $b$ and $c$
quarks~\cite{buras}. The effective Hamiltonian is written in
terms of linear combination of basis four-fermion operators:
\begin{equation}
H_{\mathrm W}^{\mathrm eff} = 
\frac{G_F}{\sqrt{2}} V_{ud}\,V^*_{us} \sum_{i=1}^{10} \Bigl[
z_i(\mu) + \tau y_i(\mu) \Bigr] O_i (\mu) 
 \, , 
\end{equation}
where $z_i$ and $y_i$ 
are Wilson coefficients (currently known at two-loop order) and
$\tau \equiv - V_{td}V_{ts}^{*}/V_{ud} V_{us}^{*}$.
In this work we seek to compute MEs $\langle\pi\pi |O_i|K\rangle$,
which are necessary for estimation of both $A_0$ and $A_2$ amplitudes 
and $\varepsilon '$. 

\begin{table}[t]
\caption{\label{tab:param}\small Lattice parameters}
\begin{tabular}{cccccc}
$N_f$ & $\beta =\frac{6}{g^2}$ & Lattice size & L, fm & Number of & 
Quark masses\\
& & & & configurations & used \\
\tableline
0 & 6.0 & $16^3\times (32\times 4)$ & 1.6 & 216 & 0.01 --- 0.05 \\
0 & 6.0 & $32^3\times (64\times 2)$ & 3.2 & 26 & 0.01 --- 0.05 \\
0 & 6.2 & $24^3\times (48\times 4)$ & 1.7 & 26 & 0.005 --- 0.03 \\
2 & 5.7 & $16^3\times (32\times 4)$ & 1.6 & 83 & 0.01 --- 0.05 \\
\end{tabular}
\end{table}

For technical reasons, two-hadron states are very difficult to put 
on the lattice~\cite{testa}. We compute $\langle\pi |O_i|K\rangle$
and $\langle 0|O_i|K\rangle$ and use chiral perturbation theory to recover
$\langle\pi\pi |O_i|K\rangle$ as follows~\cite{bernard}:
\begin{equation}
\langle \pi^+\pi^-|O_i|K^0\rangle  =
\langle \pi^+|O_i - \alpha_i O_{sub}|K^+\rangle \cdot \frac{m_K^2-m_\pi^2}
{(p_\pi\cdot p_K)f} \, , 
\end{equation}
where
\begin{equation}
\alpha_i  =  \frac{\langle 0|O_i |K^0\rangle}{\langle 0|O_{sub}|K^0\rangle} \,. 
\end{equation}
With staggered fermions, there is only one (lower-dimensional) operator 
that needs to be subtracted non-perturbatively in the manner shown above, 
namely
\begin{equation}
O_{sub} \equiv (m_d+m_s)\bar{s}d + (m_d-m_s)\bar{s}\gamma_5d \,.
\end{equation}

It should be kept in mind that this procedure is based on the lowest-order 
relationships in the
chiral perturbation theory, and thus may be significantly biased.
In particular, it ignores the final state interactions between pions
among other higher-order corrections.

\begin{figure}[ht]
\begin{center}
\leavevmode
\centerline{\epsfysize=10cm \epsfbox{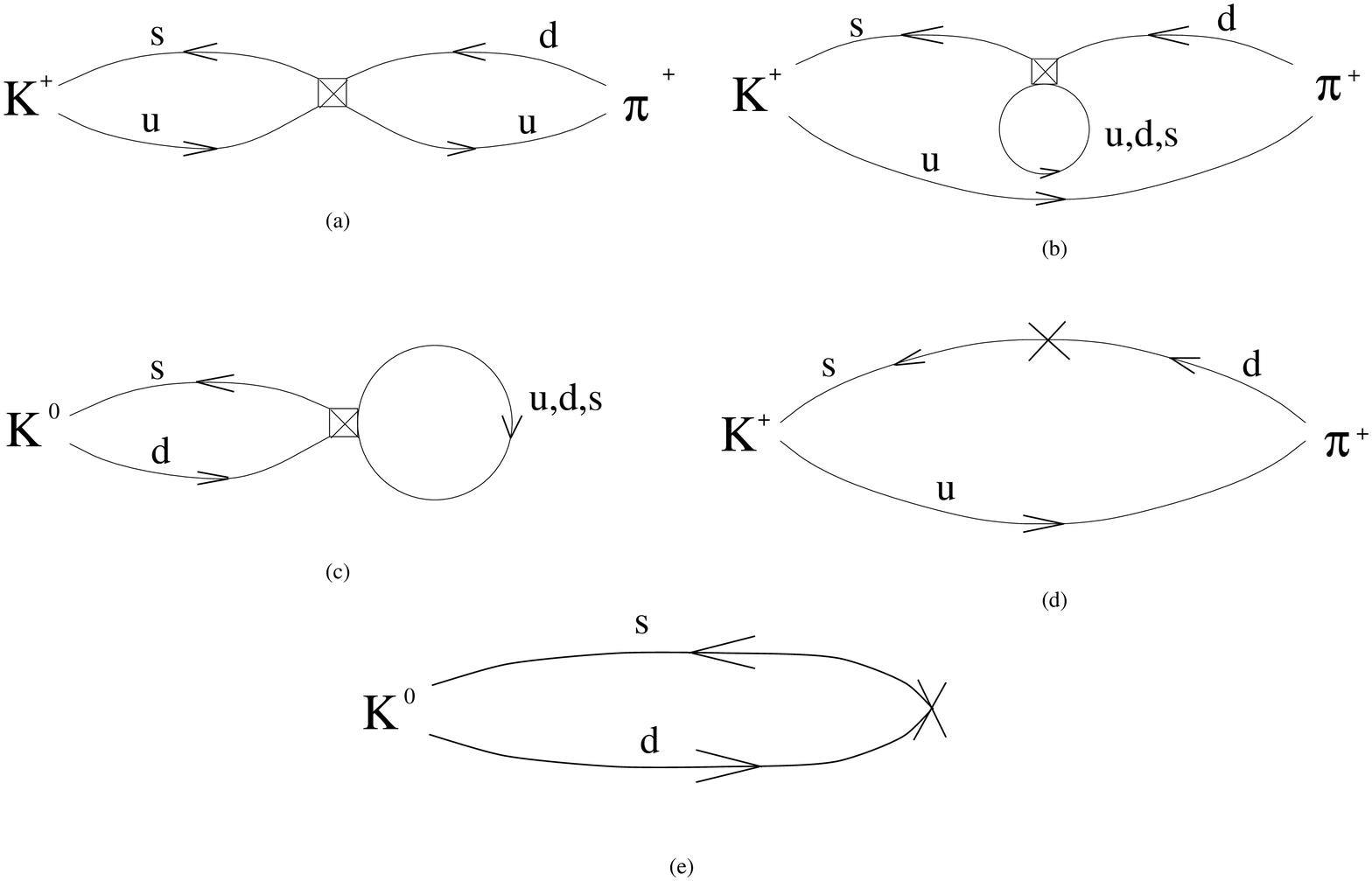}}
\end{center}
\vskip -.2cm
\caption{\label{diag}
\small Five diagrams types needed to be computed: (a) ``Eight'';
(b) ``Eye''; (c) ``Annihilation''; (d) ``Subtraction''; (e)
two-point function.}
\end{figure} 

In Table~\ref{tab:param} we show the simulation parameters. 
We are using a quenched $\beta =6.0$ ensemble, and compare
it with a quenched $\beta =6.2$ ensemble to study cutoff dependence,
with a $\beta =5.7$ ensemble with two dynamical quark flavors (with 
comparable cutoff size) to check the magnitude of quenching effects,
and with another quenched $\beta =6.0$ ensemble to check the finite
volume dependence. 

Shown in Fig.~\ref{diag} are the diagrams we are computing.
The ``eight'' diagrams, along with the two-point function and 
``subtraction'' diagrams, are relatively cheap to compute. In contrast, 
the ``eye''
and ``annihilation'' diagrams have been known in the past to be noisy and
therefore difficult to analyze. In the present study we have gained 
enough statistics to compute them with reliable accuracy.   

We use staggered fermions and gauge-invariant, tadpole-improved
operators throughout the simulation. The masses of all light quarks
are equal. Other details and exact
expressions for various quantities can be found in our
upcoming paper~\cite{PK1}. 

\section{Results of the simulation: $\Delta I=1/2$ rule}

\begin{figure}[htbp]
\begin{center}
\leavevmode
\centerline{\epsfysize=10cm \epsfbox{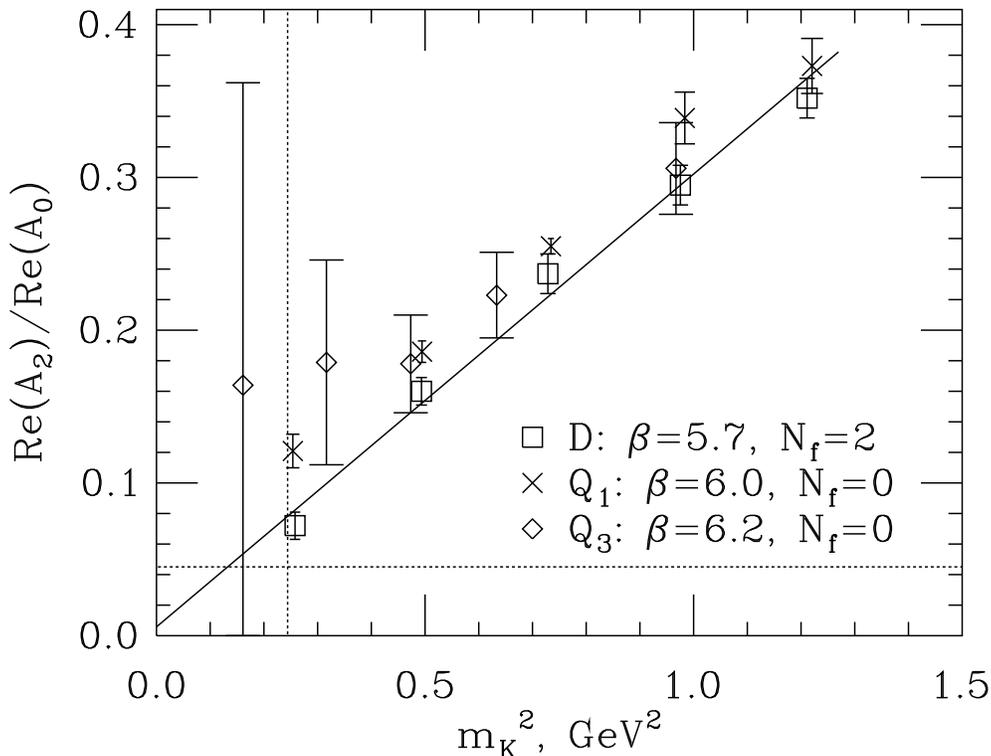}}
\end{center}
\vskip -.2cm
\caption{\label{omega}
\small The ratio Re$A_2/\mbox{Re} A_0$ versus the meson mass squared
for quenched and dynamical ensembles.
The dynamical ensemble data were used for the fit. 
The horizontal line shows the experimental
value of $1/22$. The vertical line corresponds to the physical
kaon mass. The error bars show only the statistical errors.}
\end{figure}

\begin{figure}[htbp]
\begin{center}
\leavevmode
\centerline{\epsfysize=10cm \epsfbox{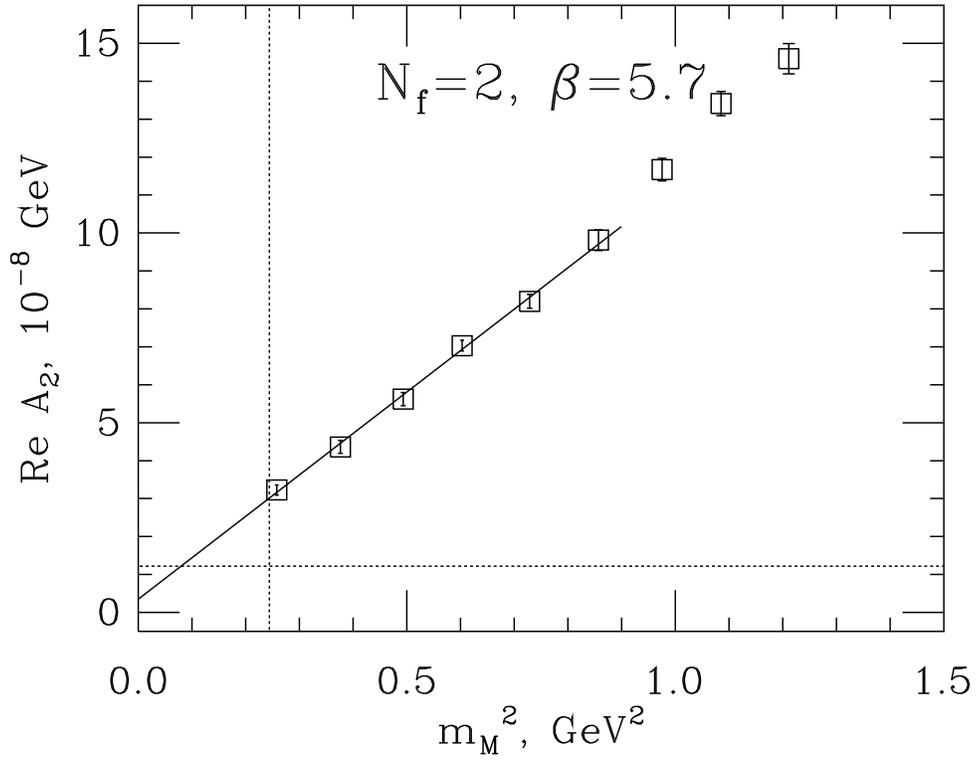}}
\end{center}
\vskip -.2cm
\caption{\label{A2}
\small Re$A_2$ for the dynamical ensemble. 
The horizontal line is the experimental value of 1.23 GeV. }
\end{figure}

\begin{figure}[htbp]
\begin{center}
\leavevmode
\centerline{\epsfysize=10cm \epsfbox{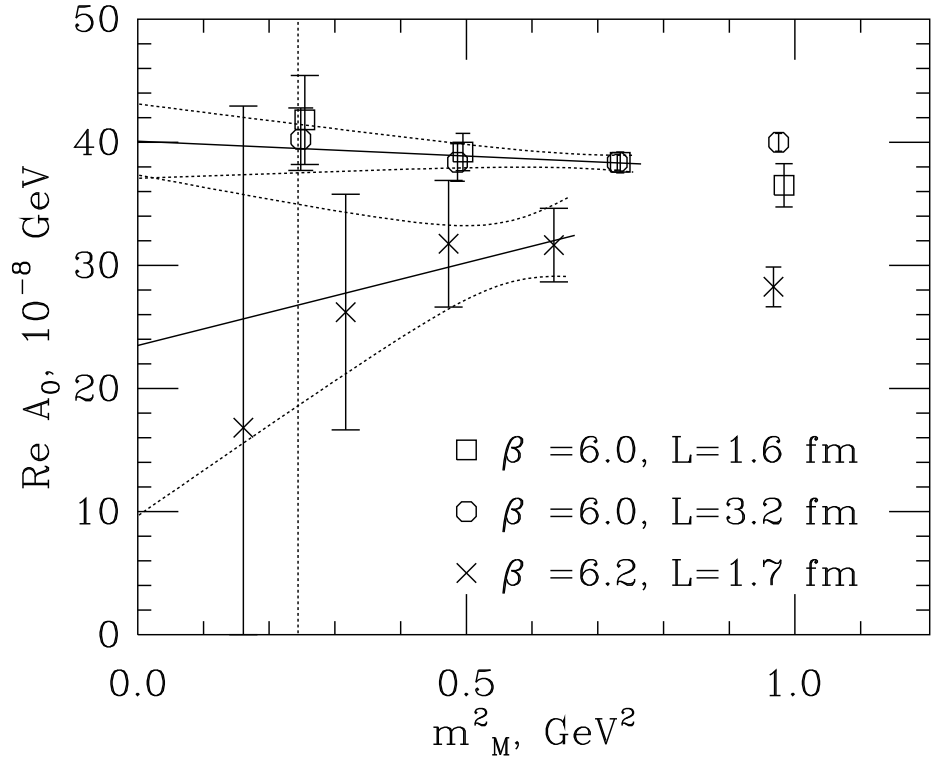}}
\end{center}
\vskip -.2cm
\caption{\label{A0}
\small Re$A_0$ for quenched ensembles plotted against the meson mass 
squared. The upper group of points
is for ensembles $Q_1$ and $Q_2$, while the lower group is for $Q_3$. 
Only statistical errors are shown. }
\end{figure}

\begin{figure}[htbp]
\begin{center}
\leavevmode
\centerline{\epsfysize=6cm \epsfbox{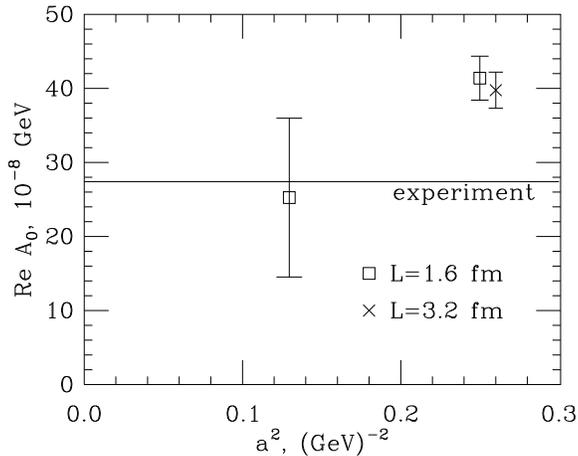}}
\end{center}
\vskip -.2cm
\caption{\label{cutoff}
\small Re$A_0$ for quenched ensembles plotted against lattice spacing
squared. The horizontal line shows the experimental result of 
$27.2\cdot 10^{-8}$ GeV. Only statistical errors are shown.}
\end{figure}

The main results of this work are shown in Fig.~\ref{omega}.
The dependence of isospin amplitude ratio on the 
kaon mass is quite dramatic. This is due to the behaviour of 
Re$A_2$ amplitude, shown in Fig.~\ref{A2} (for dynamical ensemble).
The Re$A_0$ amplitude is rather weakly dependent on the kaon mass
(Fig.~\ref{A0}). Dependence of Re$A_0$ amplitude on the cutoff
size is shown in Fig.~\ref{cutoff}. The finite volume dependence
and the effect of quenching are found small compared to noise.

In order to compare the theoretical prediction for the amplitude
ratio with experiment, it is necessary to specify a mass point
for extrapolation in Fig.~\ref{omega}. If higher-order chiral
relationships were used to obtain these amplitudes, they would
provide a natural mass point. At this stage, all we can do is
suggest that the mass point is somewhere between kaon and pion
mass. This would give us a result approximately consistent with
experiment, especially for the dynamical ensemble. Independent
of the exact mass point, a significant enhancement of $\Delta I=1/2$
over $\Delta I=3/2$ transition is evident. 

\section{Difficulties with $\varepsilon '$}

In addition to the difficulties mentioned above, estimation
of $\varepsilon '$ is further complicated by the issue of
matching lattice and continuum operators. Usually such matching
is done using lattice perturbation theory. For some operators
important for $\varepsilon '$, the perturbation theory 
is unreliable. This is a serious obstacle for estimating $\varepsilon '$,
especially considering that it is an extremely
fragile quantity, dependent on several subtle cancellations.  
What is needed is a non-perturbative matching
procedure, similar to the one described in Ref.~\cite{non-pert}.

As a temporary step, we have adopted a partially non-perturbative
matching procedure, based on computing bilinear renormalization
constants~\cite{PK1}. We can obtain only an approximate guide to the size
of four-fermion operator matching coefficients, since a few
{\it ad hoc} assumptions are made. If this procedure
is true, we get a surprising result: $\varepsilon '/\varepsilon < 0$.
This result and the validity of our assumptions should be verified by 
full non-perturbative operator matching. 

\section{Conclusion}

We have computed MEs of all basis operators appropriate for
predicting the magnitude of $\Delta I=1/2$ amplitude enhancement and 
$\varepsilon '/\varepsilon$, on a number of gauge ensembles
with reasonable statistical accuracy. A few systematic uncertainties
still preclude a definite result for either of the two quantities.
For $\Delta I=1/2$ rule, the main uncertainty is due to unknown 
higher-order chiral corrections, while for $\varepsilon '/\varepsilon$
there is an additional uncertainty due to the difficulties with 
operator matching. We observe that within this sizeable uncertainty,
the prediction for Re$A_0/\mbox{Re}A_2$ ratio
is of the same order as the experiment. 

The computations were done at the Ohio Supercomputing Center and
NERSC. We thank Columbia group for access to the dynamical configurations.

\end{document}